\documentclass[showpacs, two column]{revtex4-1}
\usepackage{graphicx}

\begin{document}

\title{Stability of $N$-soliton molecules in dispersion-managed optical fibers }

\author{Abdel\^{a}ali Boudjem\^{a}a$^{1}$ and U. Al Khawaja$^{2}$}

\affiliation{$^1$Department of Physics, Faculty of Sciences, Hassiba Benbouali University of Chlef, P.O.
Box 151, 02000, Chlef, Algeria.
\\$^2$Physics Department, United Arab Emirates University, P.O. Box 15551, Al-Ain, United Arab Emirates.}


\begin{abstract}
We investigate the stability of $N$-soliton molecules in dispersion-managed optical fibers with focus on the recently realized 2- and 3-soliton molecules. We calculate their binding energy using an averaged nonlinear Schr$\rm{\ddot o}$dinger equation. A combination of variational and numerical solutions to this equation shows that it describes well the intensity profiles and relative separations of the experimental molecules. Extending the calculation to larger values of $N$, the binding energy per soliton is found to saturate at $N\ge7$.
\end{abstract}


\maketitle


In the last two decades, developments in fiber-optic communications have demonstrated that dispersion management (DM) presents a novel attractive type of a
nonlinear carrier of information in optical fiber links. A few years ago, a stable bound state of two DM solitons in optical fibers was realized
experimentally \cite {Mitch05} and most recently three-soliton molecules in DM optical fibers were also realized by the same group \cite{Mitch}. The main
motivation behind creating such molecules is to increase the bit-rate of data transfer in optical fibers. Coding with two or more solitons per clock period
increases the alphabet beyond the binary scheme of a single soliton. In this manner, the Shannon limit \cite{shan}, which soon will be reached, may be
exceeded \cite{mitsche_pra2013}.

The main concern in soliton molecules being data carriers is their stability against disintegration. Hence,
intensive interest in their stability has emerged \cite{Malom, serhasg,usaboudj,usama, Mitch10, Turi}. The
existence of a nonzero binding energy of the soliton molecules is an indication on its stability. The
energy of a stable soliton molecule should have a minimum for a finite separation between the solitons. In
Ref.~\cite{Mitch08} it was shown that the energy of a 2-soliton molecule indeed exhibits such a minimum and
the potential of interaction was also shown to be of molecular type.

The main aim of the present work is to provide a theoretical framework that explains the stability  of 2-
and 3-soliton molecules as observed by Mitschke and co-workers \cite{Mitch,mitsche_pra2013}. Specifically,
we will show variationally that there is indeed a nonzero binding energy for 3-soliton molecules in DM
fibers. The calculation provides an estimate for the strength of the bond between the solitons and shows
regions in the parameter space where the molecule becomes unstable. Here, we address the problem of
calculating the binding energy of the soliton molecule using an averaged nonlinear Schr$\rm\ddot o$dinger
equation (NLSE). It was shown in Ref.~\cite{akira} that solitons in DM fiber can be described by an
effective nonlinear Schr$\rm{\ddot o}$dinger equation with constant coefficients and a quadratic potential.
The averaged equation is more appealing for capturing the main features of the binding mechanism. At first,
we show that the averaged equation is not integrable, hence variational and numerical approaches will be
followed. For both cases we compare the intensity profiles with the experimental ones and obtain a good
agreement. Finally, the calculation is then extended to the larger values of $N$ up to $N=12$.




We first show that the evolution of solitons in dispersion managed dissipative optical
fibers obeys a NLSE with a quadratic potential \cite{akira} which is integrable \cite{nakkeeran}. It turns
out, however, that integrability restricts the time dependence of the dispersion to the nonrealistic case
of exponential form. Therefore, an effective NLSE will be derived by averaging over one period of the
dispersion map \cite{akira}. The effective equation will then be used to calculate the binding energy of
the soliton molecule.

Solitons in dispersion-managed dissipative optical fibers are
described by the following NLSE:
\begin{equation}
i\,q_z+{d(z)\over2}\,q_{tt}+|q|^2\,q=-i\,\Gamma(z)\,q \label{gp1},
\end{equation}
where $q(t,z)$ is the envelope function of the soliton and the subscripts denote partial derivatives. Here
$z$ and $t$ are normalized distance and time , $d(z)$ corresponds to the dispersion management map defined
by
\begin{equation}
d(z)=\left\{\begin{array}{ll} d^+,\hspace{1cm}0\le z\le L^+,\\d^-,\hspace{1cm}L^+< z\le L^++L^-,
\end{array}\right.\label{doft}
\end{equation}
where $d^{+,-}$ are constant group velocity dispersions of the fiber segments $L^{+,-}$, respectively. The
loss(gain) corresponds to positive (negative) $\Gamma(z)$.

The transformation $q(t,z)=\exp{(-\int\,\Gamma(z)\,dz)}\,u(t,z)$, moves the loss term to the coefficient of
the nonlinear term:
\begin{equation}
i\,u_z+{d(z)\over2}\,u_{tt}+e^{-2\int\,\Gamma(z)\,dz}\,|u|^2\,u=0
\label{gp2}.
\end{equation}
A quadratic phase chirp develops due to the propagation of the soliton in the fiber which corresponds to
the transformation $u(t,z)=w(t,z)\,\exp{(i\,C(z)\,t^2/2)}$, where $C(z)$ is a real function. Substituting
this transformation in the last equation, the NLSE takes the form
\begin{widetext}
\begin{equation}
i\,(w_z+t\,d\,C\,w_t)+{d\over2}\,w_{tt}+e^{-2\int\,\Gamma\,dz}\,|w|^2\,w
-{1\over2}\,({\dot C}+d\,C^2)\,x^2\,w+{i\over2}\,C\,d\,w=0
\label{gp3}.
\end{equation}
\end{widetext}
With the scaling transformation $\tau=p(z)\,t$, $w(t,z)=W(t,z)\,a(z)$,
where $p(z)=\exp{(-\int{C(z)\,dz})}$ and $a(z)$ is a real function,
this equation takes the form
\begin{eqnarray}
i\,W_z&+&{d\,p^2\over2}\,W_{\tau \tau}+a^2\,e^{-2\int\,\Gamma\,dz}\,|W|^2\,W
\nonumber\\&-&{1\over2}\,{{\dot C}+d\,C^2\over
p^2}\,\tau^2\,W\nonumber\\&=&-i\,\left({\dot{a}\over
a}+{C\,d\over2}\right)\,W-i\,\left({\dot{p}\over
p}+C\,d\right)\,\tau\,W_{\tau} \label{gp5}.
\end{eqnarray}
With the choices
\begin{equation}
{\dot{a}/ a}+{C\,d/2}=0
\label{eq51},
\end{equation}
\begin{equation}
{\dot{p}/p}+C\,d=0 \label{eq52},
\end{equation} which
give $a=c_1\,\sqrt{p}$, where $c_1$ is a real arbitrary constant, and defining
\begin{equation}
\kappa={{\dot C}+d\,C^2\over p^2}
 \label{eq53},
\end{equation}
the last equation reduces to
\begin{equation}
i\,W_z+{d\,p^2\over2}\,W_{\tau \tau}+a^2\,e^{-2\int\,\Gamma\,dz}\,|W|^2\,W
-{1\over2}\,\kappa\,\tau^2\,W=0 \label{gp6}.
\end{equation}
Note that this equation can be easily put in integrable form \cite{usama_lax}. However, integrability restricts $d(z)$ to be exponential, which is not a
realistic option.
Typically, Eqs.~(\ref{eq51}) and (\ref{eq52}) are solved using the Nijhof's method \cite{nijhof},
where a close orbit in the $p-C$ plane guarantees a
unique solution for appropriate values of $\kappa^{+}$ and $\kappa^{-}$ defined as
\begin{equation}
\kappa(z)=\left\{\begin{array}{ll} \kappa^{+},\hspace{1cm}0\le z\le L^+,\\
\kappa^{-},\hspace{1cm}L^+< z\le L^++L^-.
\end{array}\right.\label{eq54}
\end{equation}
A closed orbit was indeed found using the experimental parameters of Ref.~\cite{Mitch} (see below), as shown in Fig.~\ref{fig1new}. Using the solutions for
$p(z)$ and $C(z)$, Eq.~(\ref{gp6}) can be averaged over one dispersion period to give
\begin{equation}\label{gp8}
iW_ z +\frac{\beta}{2}W_{\tau \tau} +A_0|W|^2 W-{1\over2}K_0 \tau^2\,W=0\label{gp100},
\end{equation}
where $<\dots>\equiv\int_0^L(\dots)dz/L$, $\beta=<dp^2>$, $A_0=c_1^2<p>$ and $K_0=<\kappa>$, and a dissipationless fiber, $\Gamma(z)=0$, was considered.
Thus, solitons in a dispersion-managed fiber are described by this averaged equation.
It does indeed describe the core and oscillatory tails of the
dispersion-managed solitons \cite{akira}. This equation will be used in the following
to calculate the binding energy and equilibrium size of soliton
molecules.


For numerical purposes, it is useful to reduce Eq.~(\ref{gp100}) into a dimensionless form. First we
introduce the parameters $z'=A_0\,z$, $\beta'  =\beta/A_0$ and $K'_0=K_0/A_0$. Then, Eq.~(\ref{gp100})
becomes
\begin{equation}\label{dd}
i W_{z^\prime} +\frac{\beta^\prime}{2} W_{\tau \tau} +|W|^2 W-{1\over2}\,K_0^{\prime} \tau^2W=0.
\end{equation}
We introduce the dimensionless variables $Z=z'/L'$, $T=\tau/\tau_m$ and $\Psi=W.\,\sqrt{L'}$ where $\tau_m$ is
the characteristic time scale equal to the pulse duration of the laser source and $L'=(L^{+}+L^{-})A_0$ is the length
of the dispersion map period. In terms of these parameters, the dimensionless NLSE takes the form
 \begin{equation}\label{dd}
i\Psi_Z +\frac{D}{2}\Psi_{TT} +|\Psi|^2 \Psi-{1\over2}B T^2\Psi=0,
\end{equation}
where $D=\beta^\prime L'/ \tau^2_{m}$ and $B=K'_0L'\tau_{m}^2$. We use the experimental parameters for the DM map corresponding to the setup of
\cite{Mitch}. The pulse duration $\tau_{m}=0.25$ ps, $d^{-}=-4.259\, {\rm{ps}^2/km}$, $d^{+}=5.159\, {\rm{ps}^2/km}$, $A_0=1.7\, {\rm W^{-1}km^{-1}}$, $
L^+=24$ m, $ L^-=22$ m, $L'=0.078{\rm W^{-1}}$ . Note that $d(z)$ here is the negative of that in Refs.~\cite{Mitch, mitsche_pra2013}. Using these
experimental values in the Nijhof's method, as described above, we get $\beta=<dp^2>=0.71$, $K_0=<\kappa>=-0.0156$. Notice that $A_0=c_1^2<p>=1.7$ is given
as an experimental parameter which accounts for a specific selection of the arbitrary constant $c_1$. Thus, the scaled coefficients $D$ and $B$ take the
values $D=0.521$ and $B=-4.5\times 10^{-5}$.



We use a variational calculation to show that 2- and 3-soliton molecules have indeed nonzero binding energy.
This will be evident from the minimum of the energy in terms of a finite separation between solitons.
The depth of the minimum will give an estimate to the strength of the bond in the molecule.

We employ the following three solitons trial wavefunction
 \begin{equation}\label{var1}
\Psi(Z,T)=A\sum\limits_{j=1}^ {3} A_j\exp \left[-\frac{(T-\eta_j)^2}{q^2}+i\varphi_j\right],
\end{equation}
where $A$  guarantees the normalization of $\Psi$ to the number of solitons in the molecule, namely $N=3$.
The variational parameters $q(Z)$, $\varphi (Z)$ and $\eta (Z)$ correspond respectively to the width, the
phase, and the peak position of the soliton. The energy functional corresponding to Eq.(\ref{dd}) reads
 \begin{equation}\label{enr}
E=\int_{-\infty}^\infty \left[\frac{D}{2} |\nabla\Psi|^2+B Z^2|\Psi|^2-\frac{1}{2} |\Psi|^4 \right]dT\label{efunc}.
\end{equation}
In  real units, this  energy can be expressed  as $E[J]= (\tau_m/L') E$.


We plot in Fig.~\ref{bin31} the energy as a function of the soliton width, $q$, where the experimental values of all other parameters, as in
\cite{mitsche_pra2013}, have been used. A minimum is obtained at $q=0.32$ ps which is close to the experimental value for the three-soliton molecule. In
terms of the separation between solitons, $\Delta_1=\eta_1-\eta_2$ and $\Delta_2=\eta_2-\eta_3$, Fig.~\ref{bin32} shows that the energy has a local minimum
for finite values of separations, namely at $(\Delta_1,\Delta_2)=(-1.0,-1.2)$. This agrees also with the experimental values for equilibrium separations,
as will be shown more clearly below in Fig.~\ref{mcp23}. The fact that there is also a minimum at $\Delta_1=\Delta_2=0$ indicates that the stability of the
soliton molecule is reduced due to the possibility of tunneling from the off-centered minima to the one at the origin, signifying the merging of the three
solitons into one.

The value of the energy at the local minimum represents a measure of the strength of the bond in the molecule. The binding energy of the 3-soliton molecule
relative to that of the single soliton energy, denoted by $E_{ss}$, is taken from Eq.~(\ref {enr})
and equals $E/E_{ss}=0.9$.

In Fig.~\ref{mcp23} we compare the intensity profile obtained by our varaiational calculation with the
experimental and simulation results of Ref.~\cite{Mitch} for both 2-soliton and 3-solion molecules. The
figure shows a good agreement between the variational calculation at one hand and the experimental data and
the direct numerical solution of the NLSE at the other hand. Our curves were calculated using the
experimental values of all parameters apart from the separations between solitons which were left as
variational parameters. The intensity profile was then calculated using the solitons separations obtained
by minimizing the energy functional.

To study the equilibrium properties of $N$-soliton molecules, we extend  the variational formalism of the
previous section to larger values of $N$.
Therefore, we consider the generalized Gaussian trial function
 \begin{equation}\label{var2}
\Psi(T,Z)=A \sum\limits_{j=1}^ {N_s} A_j\exp
\left[-\frac{(T-\eta_j)^2}{q^2}+i\varphi_j\right]\label{eq122},
\end{equation}
where $A_j=A_0(1+C(-1)^j)$, $\eta_j=-{(N_s-1)}/2+(j-1)\Delta$, and $\varphi_j= \pi (-1)^j$ with $N_s$
being the number of solitons in the molecule. This represents a string of $N$ solitons each of width $q$,
amplitude $A_j$, center-of-mass position $\eta_j$, and phase $\phi_j$. For simplicity, the amplitudes were
taken to alternate between $A_0(1+C)$ and $A_0(1-C)$, the separation between adjacent solitons is $\Delta$,
and phases alternating between $\pi$ and $-\pi$. In this manner, the parameter space is reduced to two
parameters only, $C$ and $\Delta$. Using the trial function, Eq.~(\ref{eq122}), in the energy functional,
Eq.~(\ref{efunc}), the energy of the 2-soliton molecule takes the form

\begin{equation} \label{E2}
E_2=\frac{{2\sqrt{\pi}q^3 D \Lambda_1 \Lambda_3}+B \Lambda_1 \Lambda_2\sqrt{\pi}\,q^4\, e^{-\frac{\Delta
^2}{2 q^2}} -{2q^3 \Lambda_4}}{4 \sqrt{\pi}q^4\Lambda_1^2},
\end{equation}
where\\
\begin{math}
\Lambda_1=\left(C ^2+1\right) e^{\frac{\Delta ^2}{2 q^2}}-\left(C ^2-1\right) \cos (2 \pi  \delta ),
\end{math}\\
\begin{math}
\Lambda_2=\left(C ^2+1\right) e^{\frac{\Delta ^2}{q^2}} \left(q^2+\Delta ^2\right)-q^2 \left(C^2-1\right) \\ \times \sqrt{e^{\frac{\Delta^2}{q^2}}} \cos (2 \pi  \delta ),
\end{math}\\
\begin{math}
\Lambda_3=q^2 \left(C ^2+1\right) e^{\frac{\Delta ^2}{2 q^2}}-\left(C ^2-1\right) \left(q^2-\Delta^2\right) \cos (2 \pi  \delta ),
\end{math}\\
\begin{math}
\Lambda_4=-4 \left(C ^4-1\right) e^{\frac{\Delta ^2}{4 q^2}} \cos (2 \pi  \delta )+\left(C ^4+6 C^2+1\right) e^{\frac{\Delta^2}{q^2}} \\+ \left(C ^2-1\right)^2 (\cos (4 \pi  \delta )+2).
\end{math}\\\\
For larger $N$ values, the energy  expressions become even more lengthy and hence will not be shown here for convenience. One can see from Fig.~\ref{mcp10}
that the binding energy of three solitons is the largest (in magnitude) among all other molecules. The binging energy starts to stabilize when the number
of solitons reaches 7. Another important remark is that the odd number of soliton molecules is more stable than the even ones. A similar behavior is
obtained for the separation where it decreases with increasing number of solitons to saturate at $N_s \ge 7$. Also, a careful observation to the same
figure shows that the separation between the second and the third solitons converges at $N_s \ge 4$  to zero, which means that the two solitons merge.


In conclusion,  we have considered $N$-soliton molecules propagating in dispersion-managed optical fibers.
The well-known effective nonlinear Schr$\rm\ddot o$dinger equation was first rederived and then shown to be
not integrable for the realistic situation. Using a variational calculation, the binding energy of the
soliton molecules was calculated with emphasis on the 2- and 3-soliton molecules. From the locations and
depths of the local minima in the equilibrium energy of the molecule, the bond length and strength were
calculated. The calculated sizes of the 2- and 3-soliton molecules agreed favorably with the experimental
values and the numerical simulation as shown in Fig.~\ref{mcp23}. It should be noted that Raman scattering
and higher order dispersion terms aught to affect the equilibrium properties of the molecule. These effects
will be investigated in a future work. The calculation was also extended for larger molecules to show a
nonmonotonic dependence of the molecule's binding energy and size in terms of the number of its
constituent solitons. \\

We are indebted to Fedor Mitschke and his co-workers for useful discussions and for providing us with the
experimental data. We acknowledge support of King Fahd University of Petroleum and Minerals under research
group projects RG1217-1 \& RG1217-2. U.K. Acknowledges the support of the UAEU-NRF 2011 research grant.

\begin{figure}[width]
  \centering
  \includegraphics[scale=0.6, angle=0]{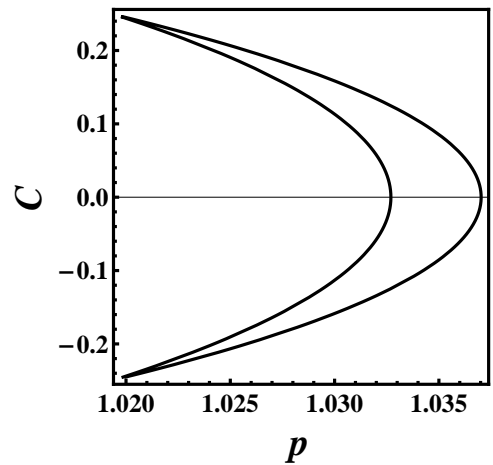}
  \caption {A closed orbit in the $p-C$ plane, which solves Eqs.~(\ref{eq51})-(\ref{eq53}), with the experimental parameters
  $d^{-}=-4.259\, {\rm{ps}^2/km}$, $d^{+}=5.159\, {\rm{ps}^2/km}$, $ L^+=22\,{\rm m}$, $L^-=24\,{\rm m}$  of Ref.~\cite{mitsche_pra2013}
  and the assignments $\kappa^{+}=0.009$ and $\kappa^{-}=-0.01$.}
  \label{fig1new}
\end{figure}

\begin{figure}[width]
  \centering
  \includegraphics[scale=0.75, angle=0]{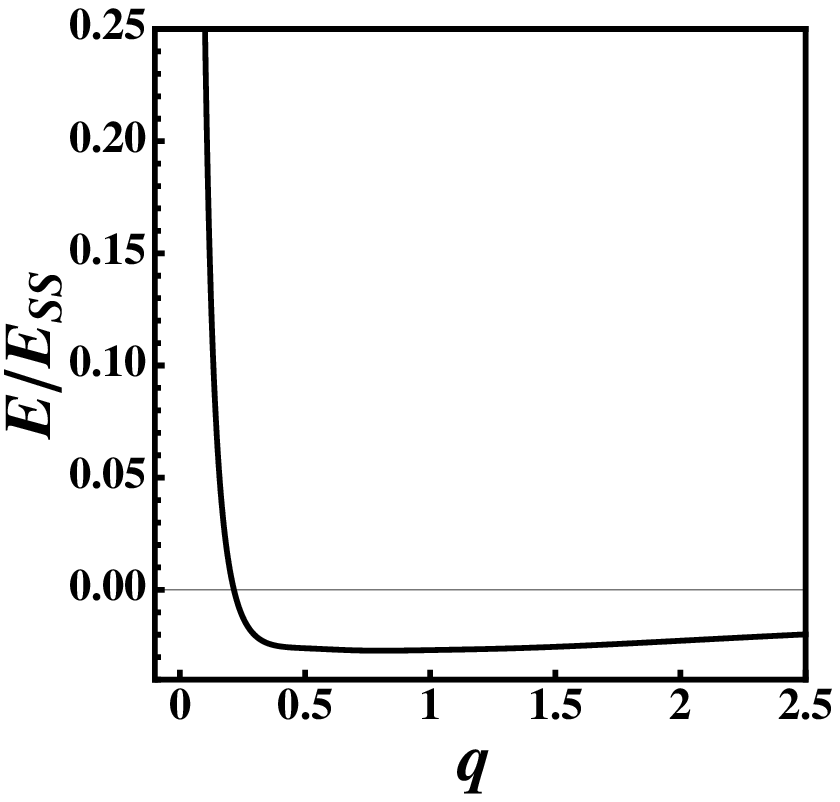}
  \caption {Binding energy of three solitons molecule relative to that of the single soliton energy $E_{ss}$ as
  function of the width $q$ with same experimental parameters of Refs.~\cite{Mitch,mitsche_pra2013} and for $D=0.521$ and $B=-4.5\times 10^{-5}$.}
  \label{bin31}
\end{figure}

\begin{figure}[bin]
  \centering
  \includegraphics[scale=0.5, angle=0]{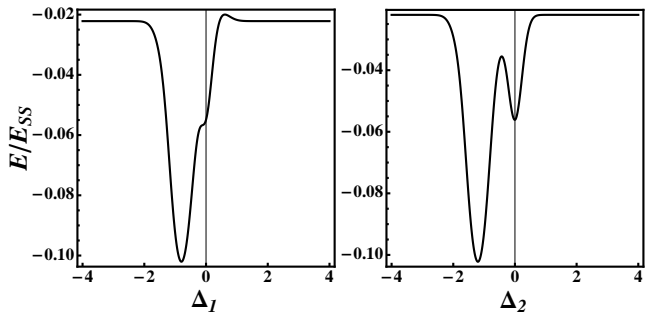}
  \caption {Binding energy of three solitons relative to that of the single soliton energy $E_{ss}$ as function of the separation $\Delta_1$ and $\Delta_2$ with same experimental parameters of Refs.~\cite{Mitch,mitsche_pra2013} and for $D=0.521$ and $B=-4.5\times 10^{-5}$.}
  \label{bin32}
\end{figure}

\begin{figure}[sol23]
  \centering
  \includegraphics[scale=0.5, angle=0]{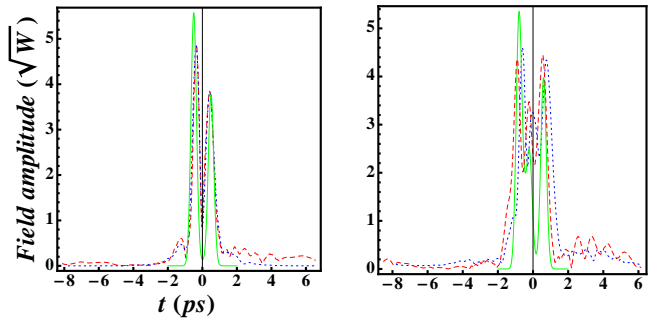}
  \caption{(Color online) Field amplitude envelopes  along dispersion managed fiber
  of two-soliton molecule (left), three-soliton molecule (right).
  Green line: our variational calculation, Blue line:
  experimental data of \cite{Mitch} and Red line: simulation of \cite{Mitch}.}
  \label{mcp23}
\end{figure}

\begin{figure}[33]
  \centering
  \includegraphics[scale=0.5, angle=0]{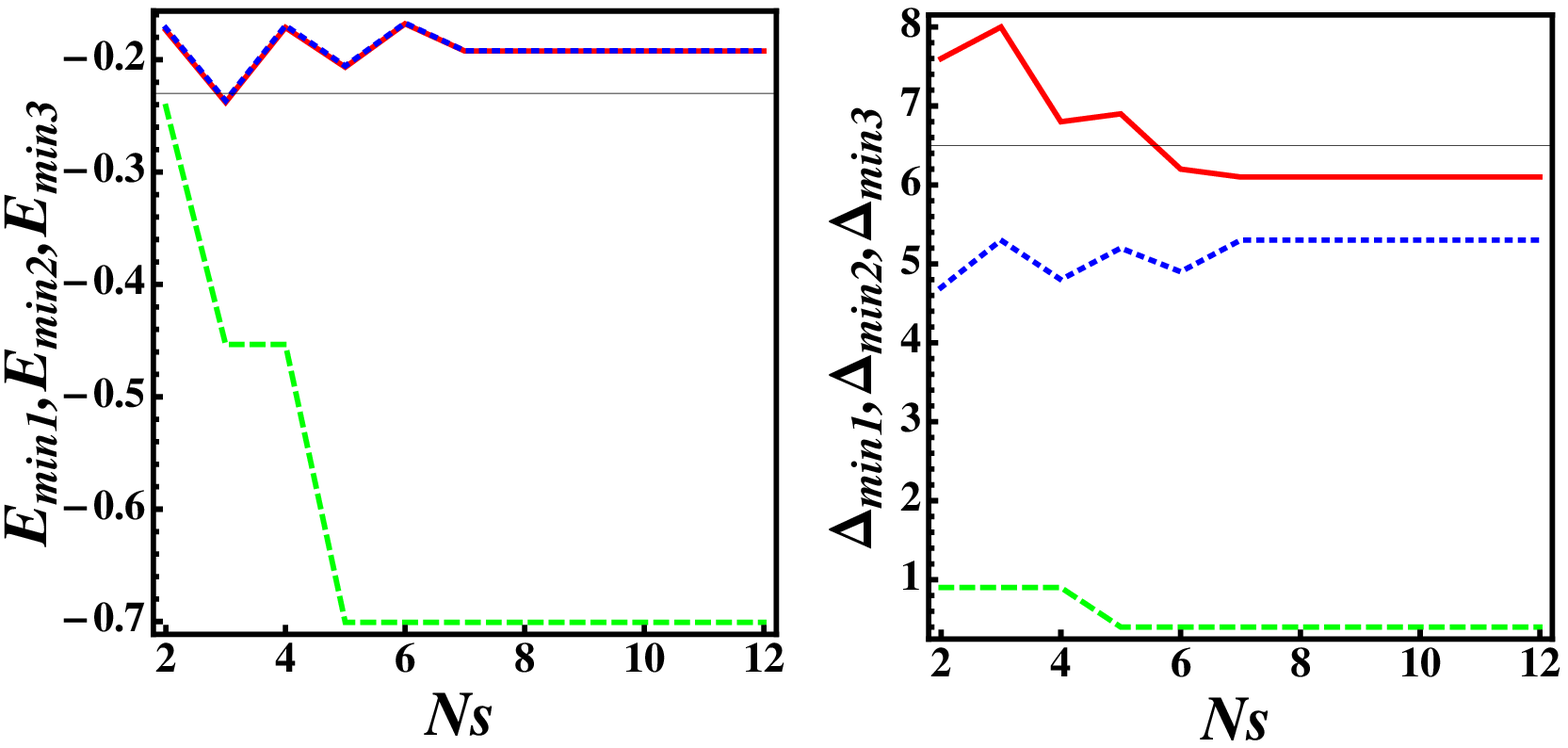}
  \caption{(Color online) Left panel: Energy minimum versus the number of solitons for $A_0=1$,
  $C=0.5$, $\delta=0.5$, $D=0.521$ and $B=-4.5\times 10^{-5}$. Red line:$E_{min1}$,
  Blue dotted line:$E_{min2}$ and Green dashed  line:$E_{min3}$. Right panel: Solitons seprations
  versus the number of solitons for same parameters. Red line:$\Delta_{min1}$,
  Blue dotted line:$\Delta_{min2}$  and Green dashed line:$\Delta_{min3}$. $E_{min1,2,3}$ and $\Delta_{min1,2,3}
  $ correspond respectively to the energy and separation at the first,
  second and the third minima of the energy.}
  \label{mcp10}
\end{figure}


\begin{thebibliography}{2}

\bibitem{Mitch05} M. Stratmann, T. Pagel, F. Mitschke, Phys.Rev.Lett {\bf95}, 14 (2005).
\bibitem{Mitch} P. Rohrmann, A. Hause, F. Mitschke, Sci.Rep. {\bf2}, 866 (2012).
\bibitem {shan} C. E. Shannon,“A Mathematical Theory of Communication”, The Bell System Technical Journal, Vol. 27, pp. 379423 and 623656, July, October, 1948.
\bibitem{mitsche_pra2013} P. Rohrmann, A. Hause, and F. Mitschke, Phys. Rev. A {\bf87}, 043834 (2013).
\bibitem{Malom} B. A. Malomed, “Variational methods in nonlinear fiber optics and related fields”, Progress in Optics 43, 69-191, (E. Wolf, editor: North Holland, Amsterdam, 2002).
\bibitem{usaboudj} U. Al Khawaja and Abdel\^{a}ali Boudjem\^{a}a, Phys. Rev. E {\bf86}, 036606 (2012).
\bibitem{usama} U. Al Khawaja, Phys. Rev. E 81, 056603 (2010).
\bibitem{Mitch10} A. Hause, H. Hartwig, F. Mitschke, Phys. Rev. A  {\bf82}, 053833 (2010).
\bibitem{Turi} S. K. Turitsyn, B. G. Bale, M. P. Fedoruk,  Phys. Rep. {\bf521}, 135203 (2012).
\bibitem{serhasg} V. N. Serkin, A. Hasegawa, and T. L. Belyaeva, Phys. Rev. Lett.98, 074102 (2007).
\bibitem{Mitch08}  A. Hause, H. Hartwig, M. B$\rm{\ddot o}$hm, F. Mitschke, Phys. Rev. A {\bf78}, 063817 (2008).
\bibitem{akira} A. Hasegawa, Y. Kodama, and A. Maruta, Optical Fiber Technology {\bf3}, 197 (1997).
\bibitem{nakkeeran} Z. Xu, L. Li, Z. Li, G. Zhou, and K. Nakkeeran, Phys. Rev. E {\bf 68}, 046605 (2003).
\bibitem{UsamaStoof} U. Al Khawaja, H T C. Stoof , R G. Hulet , K E. Strecker  and  G. B. Partridge,  Phys. Rev. Lett. {\bf89}, 200404 (2002).
\bibitem{usama_lax} U. Al Khawaja, J. Phys. A: Theor. Gen. {\bf 39}, 9679 (2006).
\bibitem{nijhof}J. H. B. Nijhof, N. J. Doran, W. Forysiak, and F. M. Knox,
Electron. Lett. {\bf33}, 1726(1997); J. H. B. Nijhof, W. Forysiak, and N. J. Doran, IEEE J. Select. Topics
Quantum Electron., {\bf6}, 330 (2000).
\bibitem{Agr} G. P. Agrawal, Nonlinear Fiber Optics, 2nd edn. (Academic Press, New York, 1995).
\end{thebibliography}
\end{document}